# Counterfactual Explanation of Brain Activity Classifiers using Image-to-Image Transfer by Generative Adversarial Network

Teppei Matsui[1,2,†,*], Masato Taki[3,†], Trung Quang Pham[4],
Junichi Chikazoe[4,5], Koji Jimura[6]

1. Department of Biology, Okayama University, Okayama, Japan
2. JST-PRESTO, Japan Science and Technology Agency, Tokyo, Japan
3. Graduate School of Artificial Intelligence and Science, Rikkyo University, Tokyo Japan
4. Supportive Center for Brain Research, National Institute for Physiological Sciences, Okazaki, Japan
5. Araya Inc., Tokyo, Japan
6. Department of Biosciences and Informatics, Keio University, Yokohama, Japan

\* Corresponding Author
† Co-1st authors

Acknowledgements
Data were provided in part by the Human Connectome Project, WU-Minn Consortium (Principal Investigators: David Van Essen and Kamil Ugurbil; 1U54MH091657) funded by the 16 NIH Institutes and Centers that support the NIH Blueprint for Neuroscience Research; and by the McDonnell Center for Systems Neuroscience at Washington University. This study was supported by JSPS Kakenhi (20H05052 and 21H0516513 to TM, 19K20390 to TQP, 19H04914 and 20K07727 to KJ, 21H02806 to JC); a grant from Japan Agency for Medical Research and Development (AMED) to JC (grant number JP21dm0207086). a grant from Brain/MINDS Beyond (AMED) to TM and MT (grant number JP20dm0307031); a grant from JST-PRESTO to TM; a grant from Narishige Neuroscience Research Foundation to TM.

# Abstract

Deep neural networks (DNNs) can accurately decode task-related information from brain activations. However, because of the nonlinearity of DNNs, it is generally difficult to explain how and why they assign certain behavioral tasks to given brain activations, either correctly or incorrectly. One of the promising approaches for explaining such a black-box system is counterfactual explanation. In this framework, the behavior of a black-box system is explained by comparing real data and realistic synthetic data that are specifically generated such that the black-box system outputs an unreal outcome. The explanation of the system's decision can be explained by directly comparing the real and synthetic data. Recently, by taking advantage of recent developments in DNN-based image-to-image translation, several studies successfully applied counterfactual explanation to image domains. In principle, the same approach could be used in fMRI data. Because fMRI datasets often contain multiple classes (e.g., multiple behavioral tasks), the image-to-image transformation applicable to counterfactual explanation needs to learn mapping among multiple classes simultaneously. Here we introduce a novel generative DNN (counterfactual activation generator, CAG) that can provide counterfactual explanations for DNN-based classifiers of brain activations. Importantly, CAG can simultaneously handle image transformation among all the seven classes in a publicly available fMRI dataset. Thus, CAG could provide counterfactual explanation of DNN-based multi-class classifiers of brain activations. Furthermore, iterative applications of CAG were able to enhance and extract subtle spatial brain activity patterns that affected the classifier's decisions. Together, these results demonstrate that the counterfactual explanation based on image-to-image transformation would be a promising approach to understand and extend the current application of DNNs in fMRI analyses.

# Introduction

Recent studies demonstrated promising results of the deep neural network (DNN) (LeCun et al., 2015) for decoding cognitive or behavioral information from brain activity images as observed with functional magnetic resonance imaging (fMRI) (Tsumura et al., 2021; Wang et al., 2020). However, despite these promising results, further applications of DNN to fMRI data could be limited due to its poor interpretability. Because of its highly non-linear and complex processing, it is often difficult to interpret what features of a given input led to the DNN's decision (Dong et al., 2019). For example, in the case of brain activity decoding, even though the DNN can accurately assign brain activations to a particular task, it is difficult to pinpoint which parts of brain activations were important for the DNN's decisions. Such interpretability would be even more important when the DNN's decoding is incorrect. Gradient-based visualization methods, such as Grad-CAM (Selvaraju et al., 2020), are frequently used to highlight image regions potentially relevant for the DNN's decision [see (Tsumura et al., 2021) for an application in neuroimaging]. However, several limitations of the gradient-based methods, such as high numbers of false positives (Eitel et al., 2019), have been reported. Thus, alongside improving the gradient-based methods (Chattopadhay et al., 2018), it would be beneficial to explore alternative approaches for interpreting the inner workings of DNNs (Adadi and Berrada, 2018).

Counterfactual explanation is one of the major approaches for explaining DNN's inner working (Goyal et al., 2019; Wang and Vasconcelos, 2020). To explain how the decision on a given data was made, counterfactual explanation uses artificial data ("counterfactuals") that are generated from the real data but targeted to an unreal outcome (decision). By comparing the DNN's decision on the real data and the counterfactual, one can deduce explanations of the decisions made by the DNN. For example, consider a case in which a brain activity classifier incorrectly assigns a gambling task to a brain activation produced in a motor task (Fig. 1a). We consider a minimal transformation of the original brain activation to a counterfactual activation that is classified (correctly) as a motor task activation by the DNN classifier (Fig. 1b). By directly comparing the original brain activation and the counterfactual activation, one can explain the classifier's decision by making a statement such as, "This brain activation map would have been correctly classified to the gambling task if brain area X had been activated." (Fig. 1c). As in this example, counterfactual explanation can provide intuitive explanations of a black-box decision system without opening the black-box, which is a critical aspect of the technique.

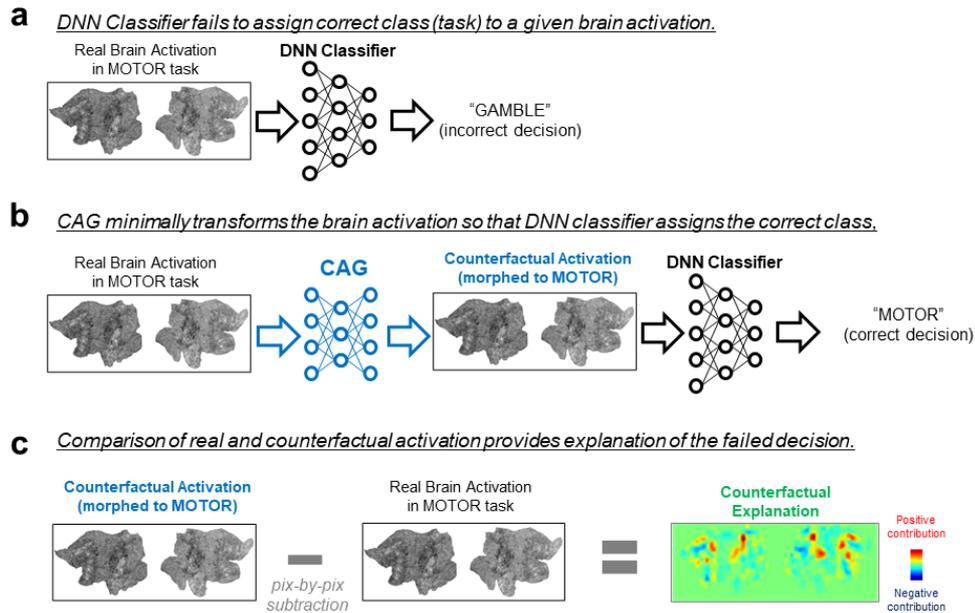

**Figure 1. Applications of counterfactual explanation in fMRI.** The example illustrates an application of counterfactual explanation to a misclassification by a DNN classifier. **a)** In this example, a DNN classifier incorrectly assigned EMOTION to a map of brain activation obtained in a MOTOR task. Because of the black-box nature of the DNN classifier, it is difficult to explain why the misclassification occurred. **b)** A generative neural network for counterfactual brain activation (CAG) minimally transforms the real brain activation in (a) so that the DNN classifier now assigns MOTOR to the morphed activation (counterfactual activation). **c)** Counterfactual explanation of misclassification in (a) can be obtained by taking the difference between the real activation and the counterfactual activation. In this example, the real brain activation would have been classified (correctly) as MOTOR if red (blue) brain regions in the counterfactual explanation had been more (less) active.

Although the generation of counterfactuals for high-dimensional data such as natural images and medical images had been difficult, recent advancement in DNN-based image-generation has made counterfactual explanation applicable to these domains. For natural images, several studies have successfully used counterfactual explanation to explain the behavior of DNN-based image classifiers (Chang et al., 2019; Liu et al., 2019; Singla et al., 2020; Zhao, 2020). In medical image analyses, counterfactual explanation has also been applied to DNN-based classifiers of X-ray and structural MR images (Mertes et al., 2020; Pawlowski et al., 2020). However, to the best of our knowledge, counterfactual explanation has not been utilized for DNN-based classifiers of fMRI data.

In the present study, we developed a generative neural network named

Counterfactual Activation Generator (CAG), which provides counterfactual explanations for a DNN-based classifier of brain activations observed with fMRI. This study aims to provide a proof of principle that counterfactual explanation can be applied to fMRI data and the DNN-based classifiers of brain activations. We demonstrated several applications of CAG. First, CAG could provide counterfactual explanations of correct classifications of the brain activations by DNN-based classifiers. Specifically, the counterfactual explanation highlighted the patterns of brain activations that were critical for the DNN classifier to assign the brain activations to particular tasks. Similarly, CAG could provide counterfactual explanations of incorrect classifications by DNN-based classifiers. Moreover, iterative application of CAG accentuated and extracted subtle image patterns in brain activations that could strongly affect the classifier's decisions. These results suggest that the image-transfer-based methods, such as CAG, would be a powerful approach for interpreting and extending DNN-based fMRI analyses.

# Materials & Methods

*Datasets*

Training data were single-subject second level z-maps obtained during the performance of seven behavioral tasks from the S1200 release of the Human Connectome Project (N = 992; HCP; http://www.humanconnectomeproject.org/) (Barch et al., 2013; Glasser et al., 2016; Van Essen et al., 2013). From each participant, statistical z-maps were obtained for activation contrasts for the Emotional processing task (Face vs. Shape), the Gambling task (Reward vs. Loss), the Language processing task (Story vs. Math), the Motor task (average of all motions), the Relational processing task (Relational processing vs. Matching), the Social cognition task (Mental vs. Random), and the N-back working memory task (2-back vs. 0-back). For brevity, the seven tasks are denoted as follows: 1) EMOTION, 2) GAMBLING, 3) LANGUAGE, 4) MOTOR, 5) RELATIONAL, 6) SOCIAL, and 7) WM (working memory). We used gray-scaled flat 2D cortical maps (Glasser et al., 2016) provided from HCP for dimensional compatibility of images between VGG16-ImageNet and activation maps. The flattened maps were created using the Connectome Workbench (https://www.humanconnectome.org/software/connectome-workbench/) following a procedure described in Tsumura et al., 2021 (Fig. 2a).

*DNN classifier of brain activations*

The DNN classifier of brain activations used in this study was adapted from our previous study (Tsumura et al., 2021) (Fig. 2b). Briefly, the DNN classifier used VGG16 pretrained on ImageNet (Simonyan and Zisserman, 2015) as the base. The DNN classifier was then trained using transfer learning to classify individual brain activity images to the seven behavioral tasks (Pan and Yang, 2010). To enable processing by generative neural networks described below, activation maps were spatially down-sampled from 570 by 1320 pixels to 50 by 140 pixels. Data were split into training data (N = 4730) and validation data (N = 518) (note that some participants in the dataset did not complete all seven tasks). Training was conducted using the training data with ten-fold cross-validation. Hyperparameters for the training were as follows: batch size, 10; epoch, 50; learning rate, 0.0001; optimizer, stochastic gradient descent (SGD); loss function, categorical cross-entropy. Pixels outside of the brain were set to zero. Model training and testing were implemented using Keras (https://keras.io/) under a Tensorflow backend (https://www.tensorflow.org/). Five instances of the DNN classifier were trained for replication. All parameters, including the training data, were the same for all the replicates.

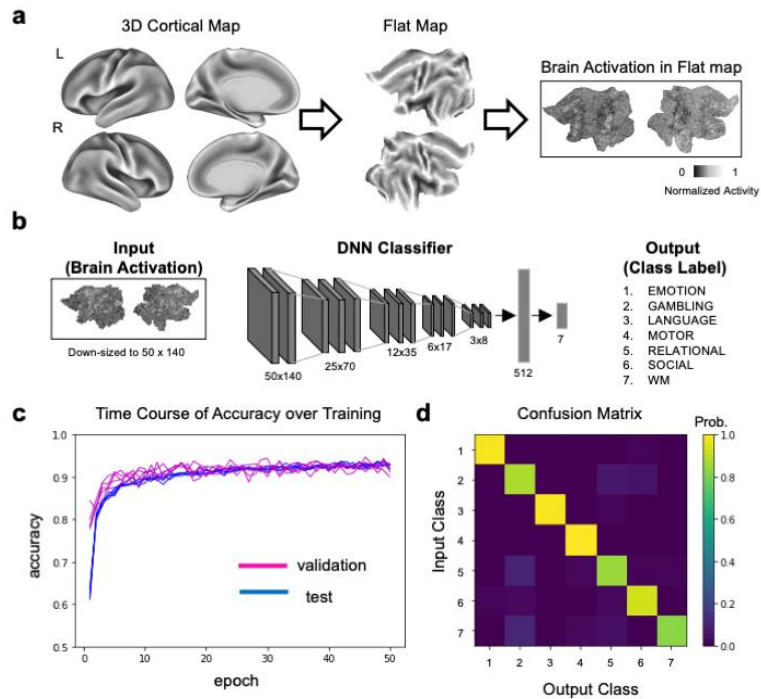

**Figure 2. DNN classifier for brain activity decoding. a)** Following the standard procedure developed by HCP (Glasser et al., 2016), neocortex in the two hemispheres were mapped to two cortical sheets. Each neocortical activity image was mapped to the two sheets, which was then input to the DNN classifier (for details, see Tsumura et al., 2021). **b)** Model architecture of the DNN classifier. The input was a picture containing two sheets of cortical activations. The picture was down-sampled for later processing by the generative neural network. The DNN classifier was a deep convolutional network similar to the one described in our previous study (Tsumura et al., 2021). The output of the DNN classifier was one-hot vectors representing seven behavioral tasks in the HCP dataset. **c)** Training history of the transfer learning. Test accuracy (blue) and validation accuracy (magenta) are shown for five replicates. Note that the chance level is 14.3% (1/7). **d)** Profile of the classifier's decision (confusion matrix) in the validation set.

*Counterfactual Activation Generator (CAG)*

We adopted the architecture of StarGAN (Choi et al., 2018), consisting of Discriminator and Generator, with some modifications (Fig 3a; see also Fig. S1 for an illustration of our overall approach). Briefly, the goal was to train a single Generator that learns mapping among multiple classes (in this case, the seven HCP tasks). We regarded this Generator as CAG. To achieve this, we trained CAG to transform a brain activation $x$ with class-label $y$ (source class) to a perturbation toward $y^c$ (target class), such that $CAG(x, y^c) \to x^c$. An auxiliary discriminator was introduced to allow a single discriminator to control multiple classes. Thus, the Discriminator produced the probability distributions over both the source and the target classes, $D: x \to \{D_{src}(x), D_{cls}(x)\}$.

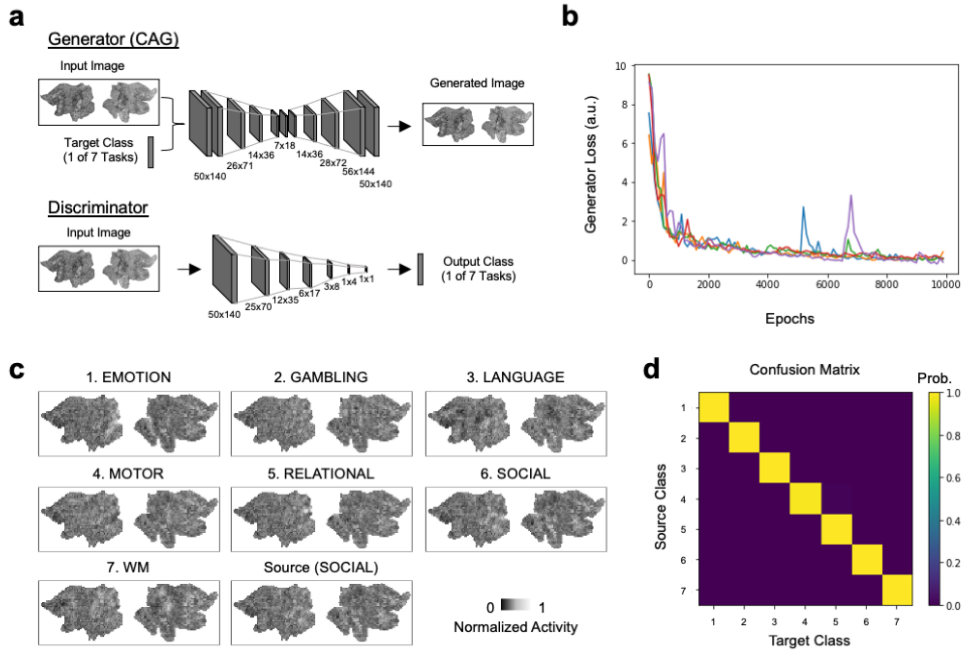

**Figure 3. Counterfactual Activation Generator (CAG). a)** Generator and Discriminator architectures. See Methods for details. Networks were modified from STAR-GAN. Generator, once trained, served as CAG. Generator takes a combination of an image of brain activation and a one-hot label indicating the target class as an input. Generator outputs a counterfactual brain activation that is a minimal transform of the input brain activation toward the target class. Discriminator takes an activation map output by Generator and outputs a one-hot label. Discriminator was co-trained with Generator, as in STAR-GAN. **b)** Time courses of the Generator Loss. Different colors indicate different replicates (n = 5). **c)** Representative counterfactual activations generated by CAG. All counterfactual activations were generated from the source activation. **d)** Confusion matrix showing the classifier's decision profile on counterfactual activations.

The loss terms were described as follows. Wasserstein Loss ($\mathcal{L}_{wass}$) and Gradient Penalty Loss ($\mathcal{L}_{gp}$) were included to make the generated brain activations indistinguishable from the real brain activations. $\mathcal{L}_{wass}$ between the real and counterfactual activations and $\mathcal{L}_{gp}$ were defined as follows,

$$\mathcal{L}_{wass} = \mathbb{E}_x[D_{src}(x)] - \mathbb{E}_{x,c}[D_{src}(CAG(x,c))]$$

$$\mathcal{L}_{gp} = \mathbb{E}_{\hat{x}}[(\|\nabla_{\hat{x}} D_{src}(\hat{x})\|_2 - 1)^2]$$

where $\hat{x}$ was sampled uniformly along a straight line between a pair of a real and a generated activation.

Domain classification loss was included to ensure that the transformed activation was properly classified as the target class. We considered two types of objectives. The first one is a domain classification loss of real activations used to optimize Discriminator ($\mathcal{L}_{cls}^r = \mathbb{E}_{x,c'}[-\log D_{src}(c'|x)]$). The second one is a domain classification loss of fake activations used to optimize CAG ($\mathcal{L}_{cls}^f = \mathbb{E}_{x,c}[-\log D_{src}(c|CAG(x,c))]$). This loss term forced CAG to generate activations that could be classified as the target classes.

Reconstruction loss ($\mathcal{L}_{rec}$) was defined using the cycle consistency loss (Kim et al., 2017; Zhu et al., 2017) as,

$$\mathcal{L}_{rec} = \mathbb{E}_{x,c,c'}[(\|x - CAG(CAG(x,c),c')\|_1)]$$

where CAG tried to reconstruct the original activation from the transformed activation.

Additionally, we included a loss term for the DNN classifier ($\mathcal{L}_{cnn}$) to force the mappings learned by CAG to be aligned with the classifier's decisions. This loss term was calculated using categorical cross-entropy over the fake activations.

The total losses for Discriminator ($\mathcal{L}_D$) and the CAG Loss ($\mathcal{L}_G$) were defined using the loss terms as follows,

$$\mathcal{L}_D = \mathcal{L}_{wass}^r + \mathcal{L}_{wass}^f + \lambda_{gp}\mathcal{L}_{gp} + \lambda_{cls}\mathcal{L}_{cls}^r$$
$$\mathcal{L}_G = \mathcal{L}_{wass} + \lambda_{cls}\mathcal{L}_{cls}^f + \lambda_{rec}\mathcal{L}_{rec} + \lambda_{cnn}\mathcal{L}_{cnn}$$

where $\mathcal{L}_{wass}^r$ and $\mathcal{L}_{wass}^f$ stand for Wasserstein loss for real and fake activations, respectively. We used $\lambda_{gp} = 10$, $\lambda_{cls} = 1$, $\lambda_{rec} = 10$, and $\lambda_{cnn} = 1$ for all experiments. All models were trained using Adam (Kingma and Ba, 2014), with $\beta_1 = 0.5$ and $\beta_2 = 0.999$. Training was done using the training data with ten-fold cross-validation for 10,000 epochs. Batch size and learning rate was set to 16 and 0.0001, respectively, in all experiments. The code for CAG will be made publicly available upon acceptance of the paper.

*Counterfactual explanation of correctly and incorrectly classified images*

Counterfactual explanation of correctly classified images was performed on the correctly classified brain activations (N = 478 out of 518 that were not used in the classifier training). Each counterfactual explanation was set to explain "Why this activation was correctly classified as class (task) A instead of class B?" To do this, the original brain activation was transformed by CAG toward class B. Counterfactual explanation was obtained by pixel-by-pixel subtraction of the original activation from the counterfactual activation. As for counterfactual explanation of correct classifications, counterfactual activations were obtained by transforming the correctly classified activations to one of the randomly chosen incorrect classes.

To quantitatively evaluate the effectiveness of counterfactual explanations, we conducted two analyses. In the first analysis, we perturbed image transformation by CAG at various levels and examined its effect on the classifier's decisions. For the perturbation, pixels in each counterfactual explanation whose values were below a chosen percentile threshold (α) were set to zero ($CE_\alpha$). Then the perturbed counterfactual explanation was added back to the original activation ($Activation_{original}$) as follows,

$$Activation_{new} = Activation_{original} + CE_\alpha$$

The resulting activation ($Activation_{new}$) was normalized to have minimum and maximum values of zero and one, respectively, and then input to the DNN-classifier. The percentile threshold (α) took values ranging from 0% to 100% with a 20% step. Note that $Activation_{new}$ is equal to the counterfactual activation and $Activation_{original}$ when α equals 0% and 100%, respectively. In the second analysis, each counterfactual explanation was compared with a "control explanation," which was calculated as the difference between the true class's average activations and the target class used for the transformation ($\Delta Ave$). The control explanation was added to the original activation ($Activation_{original}$) as follows,

$$Activation_{new} = Activation_{original} + \Delta Ave \times \kappa$$

The resulting activation ($Activation_{new}$) was normalized to have minimum and maximum values of zero and one, respectively, and then input to the DNN-classifier. The parameter for mixing (κ) took values ranging from 0 to 5 at with a 0.1 step, and was adjusted individually for each control explanation to maximize the total number of cases classified to the target class used for transformation.

Counterfactual explanation of incorrectly classified images was performed similarly on each incorrectly classified brain activation (N = 40). Each counterfactual explanation was set to explain "Why this activation was incorrectly classified as class (task) B instead of class A?" To do this, the original brain activation was transformed by CAG toward the true class A. Counterfactual explanation was obtained by pixel-by-pixel subtraction of the original activation from the counterfactual activation. The two quantitative analyses for the counterfactual explanation of correct classifications were similarly applied to the counterfactual explanation of incorrect classifications. In these analyses, the target class for the image transformation by CAG was set to the correct classes (instead of randomly chosen classes in the case of correct classification).

*Counterfactual exaggeration and feature extraction*

Counterfactual exaggeration (Singla et al., 2020) was performed by iteratively transforming a real brain activation toward one class. We performed up to eight iterations. Feature extraction was done by subtracting the third iteration from the eighth iteration. To quantitatively evaluate the extracted feature, the feature was added to each activation in the validation set (N = 518), then the summed image was input to the DNN classifier. For 12 extracted features from randomly chosen activations, the percent of activations assigned to the added feature's class were calculated.

# Results & Discussion

*DNN classifier decoded task information from brain activity with high accuracy*

We first trained a DNN classifier that was used as the target for counterfactual explanation. Brain activations were converted to flattened maps, which were then input to the DNN classifier (Fig. 2a). The DNN classifier was based on VGG16 pre-trained on the ImageNet dataset (Tsumura et al., 2021) (Fig. 2b). The pre-trained DNN classifier was trained to classify brain activation maps using transfer learning (Pan and Yang, 2010). After 50 epochs of training, the DNN classifier reached approximately 92% of classification accuracy for the held-out validation data. Similar results were obtained for a total of five replicates, suggesting high reproducibility (Fig. 2c). Figure 2d shows the confusion matrix showing the classifier's decision profile (see also Table S1 for exact values). Similar confusion matrices were obtained for all the replicates (data not shown). These results suggest that DNN classifiers could accurately decode task information from individual brain activations.

*CAG generated counterfactual activations were realistic and fooled the classifiers*

We next trained a generative neural network (CAG) for counterfactual explanations of the DNN classifier's decisions. For this, we adopted, with modifications, the architecture of StarGAN (Choi et al., 2018) that can perform image-to-image transformation among multiple classes. Two DNNs, generator (CAG) and discriminator, were simultaneously trained (Fig. 3a; Fig. S1). By including the classification loss by the DNN classifier, CAG was trained to simultaneously fool both the discriminator and the DNN classifier (Fig. S1; see Methods for details). Throughout the training, the Generator Loss, which is a good indicator of the quality of the generated image (Arjovsky et al., 2017), consistently decreased toward zero and plateaued around 10,000 epochs of training (data for five replicates are shown in Fig. 3b; see also Fig. S2 for time courses of all the loss terms). After the training, CAG could transform a real brain activation into a counterfactual brain activation that was visually indistinguishable from the real activations (Fig. 3c). Importantly, a single CAG was able to direct the transformation to any of the seven classes. The DNN classifier assigned the targeted class to the counterfactual activations at almost 100% accuracy (Fig. 3d; Table 1). Thus, these results suggest that CAG fulfilled the goal of generating counterfactual brain activations that were not only visually realistic but also fooled the DNN classifier.

| CLASS | EMOTION | GAMBLING | LANGUAGE | MOTOR | RELATIONAL | SOCIAL | WM |
|---|---|---|---|---|---|---|---|
| N correct | 518 | 518 | 518 | 512 | 518 | 518 | 518 |
| (% correct) | (100%) | (100%) | (100%) | (98.8%) | (100%) | (100%) | (100%) |

**Table 1. Decision profile of DNN classifier on counterfactual activations.** Each image in the validation set (N = 518) was morphed toward one of the seven classes and then input to the DNN classifier.

*Counterfactual explanation of misclassification by DNN classifiers*

Using CAG, we first conducted counterfactual explanation of the classifier's correct decisions. Specifically, we tried to visualize the pattern of brain activation that led the classifier to assign the correct class but not another (incorrect) class (Fig. 4a). In the first example, brain activations correctly classified as MOTOR by the DNN classifier were examined (Fig. 4b). We asked why these activations were not classified as EMOTION. To see this, a counterfactual activation was created by transforming each original activation toward EMOTION using CAG (Fig. 4c). Then, the counterfactual explanation was obtained by taking the difference between the original and the counterfactual activations (Fig. 4d). The positive and negative regions in the counterfactual explanation were the regions that had positive and negative influence, respectively, on the classifier's decision of assigning EMOTION but not MOTION to the counterfactual activation. In other words, the DNN classifier would have classified the original activation as EMOTION if the positive regions in the counterfactual explanation had been more active (and the opposite for the negative regions).

To quantitatively evaluate the counterfactual explanation, we compared it against a difference between the population-averaged activations of EMOTION and MOTOR (average of 74 and 75 activations, respectively) (Fig. 4e). Whereas the difference of average maps highlighted a small portion of brain areas in the occipital cortex, counterfactual explanation additionally found lateral temporal areas to be relevant. Because lateral temporal areas are known to be activated by emotional and facial processing (Glasser et al., 2016; White et al., 2014), it is reasonable to find these areas highlighted in the counterfactual explanation. The reason that only the occipital cortex was highlighted in the difference of average maps is most likely due to very high activation in this area in the average map of EMOTION compared to the average map of MOTOR. These differences between the counterfactual explanation and the explanation by the difference of averages can be understood as the difference between univariate and multivariate analyses (Jimura and Poldrack, 2012). The average map that is derived from the univariate analysis (pixel-based GLM) is affected by the choice of particular control conditions, thus the explanation by the difference of the averages would be affected by difference in the control conditions. In contrast, the counterfactual explanation can robustly detect relevant activations using spatial patterns of multiple pixels. Consistent with this idea, the counterfactual explanation, but not the difference of average maps, successfully highlighted the orbitofrontal areas implicated for emotional processing (Fig.4d) (Goodkind et al., 2012; Rolls et al., 2020).

The second example shows counterfactual explanation of why WM activations were not classified as LANGUAGE (Fig. 4f-i). In this case, unlike the previous example, the difference of averages of WM and LANGUAGE highlighted large portions of the brain (red regions in Figure 4i; average of 75 and 73 maps, respectively, for WM and LANGUAGE). With such large and distributed areas being highlighted, it is difficult to pinpoint particular areas without arbitrary thresholding. In contrast, the counterfactual explanation highlighted distributed but much more localized brain areas (Fig. 4h). It is evident from the counterfactual explanation that activations in frontal and temporal brain areas would have been necessary to shift the DNN classifier's decision from WM to LANGUAGE.

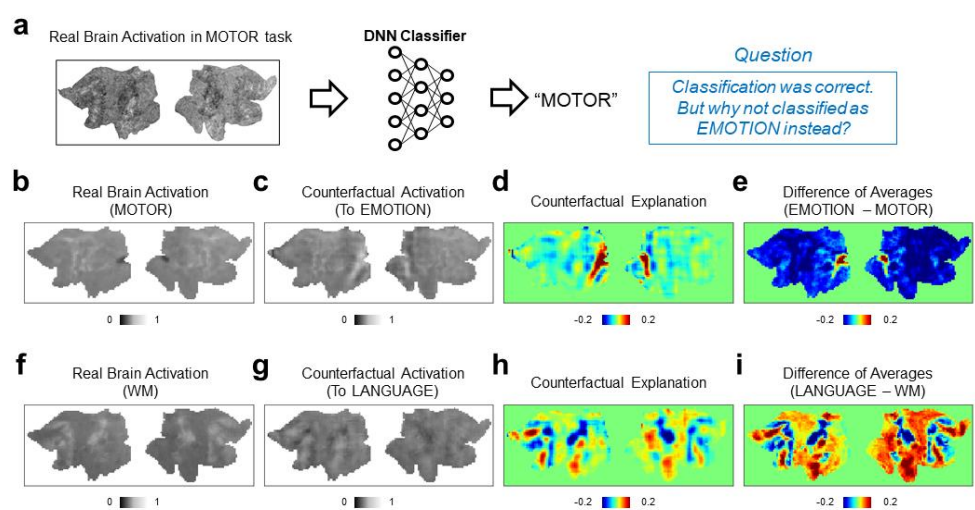

**Figure 4. Counterfactual explanation of correct classification. a)** Schematics of the question asked in this analysis. In this example, the DNN classifier correctly assigned a "MOTOR" label to a real brain activation in the MOTOR task. Here, we want to interrogate this correct decision. Specifically, we ask a question "why did the classifier assign MOTOR instead of EMOTION?". **b-e)** Examples of counterfactual explanation. (b) A population average map of real brain activation in the MOTOR task. (c) A population average map of counterfactual activation obtained by transforming the map in (a) to EMOTION. Transformation was conducted for each activation map and then averaged across the population. (d) Pixel-by-pixel subtraction of maps in (b) and (a) that serves as counterfactual explanation. This map explains why the map was classified as MOTOR but not EMOTION. (e) Simple difference between the average of real activations in the EMOTION and MOTOR tasks. **f-i)** Another example of counterfactual explanation. Same convention as in (b)-(e). In this case, we asked why brain activations were (correctly) classified as WM but not LANGUAGE.

Out of 478 correctly classified validation data, 476 counterfactual activations were classified as the targeted class. To test the robustness of the result against image corruption, we isolated the image components added by CAG (i.e., the difference between the counterfactual activation and the raw activation). Then, we perturbed the image components at different levels of percentile thresholds ($\alpha$ in Table 2. See methods) that were in turn added back to the raw activation. The effect of thresholding did not change the classification results when the bottom 20% of the image components were perturbed. The classification results were still above 25%, even when the bottom 60% of the image components were perturbed. The classification results were markedly degraded when the bottom 80% of the image components were perturbed. Thus, these results suggest that image modifications imposed by CAG were robust to perturbation in a large margin. To further assess the effectiveness of counterfactual explanation, we compared the classifier's response to counterfactual activations and control maps obtained by adding the original activation and the difference of average activations ("Control" in Table 2). Only four of the control maps were classified as the targeted class. Together, these results demonstrated that counterfactual explanations provided interpretable activation patterns that could not only explain the classifier's decisions but also robustly manipulate the classifier's decisions.

Note that the aim of the discussion here is not to infer cognitive tasks associated with the brain activation, a type of discussion considered as reverse inference (Poldrack, 2006). In this case, the cognitive tasks (i.e., classes) associated with the brain activations were entirely determined by the DNN classifier. The purpose of the discussion here is to interpret the counterfactual explanation in relation to existing knowledges about the brain activity. In the future, this type of discussion may be automated using applications such as Neurosynth (Yarkoni et al., 2011).

| Correct Cases | Counterfactual Activations | | | | | | Control |
| --- | --- | --- | --- | --- | --- | --- | --- |
| | $\alpha = 0$ | $\alpha = 20$ | $\alpha = 40$ | $\alpha = 60$ | $\alpha = 80$ | $\alpha = 100$ | |
| $N_{correct}$ (%) | 476 (99.6%) | 470 (98.0%) | 327 (68.4%) | 125 (26.2%) | 9 (1.9%) | 0 (0.0%) | 4 (0.8%) |

**Table 2. Decision of DNN classifier on counterfactual activations obtained from correctly classified brain activations.** Counterfactual activations were obtained from images correctly classified by the DNN classifier (N = 478). Each image was transformed to one of a number of randomly chosen incorrect classes. All counterfactual activations were classified as the targeted class by the DNN classifier (middle column). As for the control analysis, the difference of the average maps for targeted versus original classes was added to each image. None of the control images were classified as the targeted class (right column).

*Counterfactual explanation of misclassification by DNN classifiers*

An important feature of counterfactual explanation is its ability to provide explanations to single cases of misclassification. We next demonstrated this in misclassifications by the DNN classifier (Fig. 5a). For each case of misclassifications, the misclassified activation map was transformed toward the correct class by CAG. Then the difference between the counterfactual activation and the real (misclassified) activation was calculated for the counterfactual explanation. In the first example, a brain activation in the EMOTION task was incorrectly classified as SOCIAL (Fig. 5b). A counterfactual activation was obtained by transforming the real activation toward the correct class (EMOTION) (Fig. 5c). Interestingly, the counterfactual explanation suggested that activations in the occipital regions were critically lacking for the DNN classifier to classify the original activation as EMOTION (Fig. 5d). Because the occipital area is considered to process low-level visual information (Yamins et al., 2014), this occipital activation likely indicates bias in the dataset that used visual stimulus in the EMOTION task (Barch et al., 2013) rather than a brain activation related to emotional processing. Thus, counterfactual explanation revealed that this misclassification was likely due to the bias in the dataset, which was unintentionally learned by the DNN classifier.

As for a control analysis that can be compare with the counterfactual explanation, we calculated the difference between the misclassified (real) activation and the average activation of EMOTION (Fig. 5e). Despite the similar global trend with the counterfactual explanation, the difference with the average showed a noisy pattern whose local peaks were difficult to find. Importantly, a peak in the occipital area was difficult to discern in the difference with the average. In the second example, we examined an activation in WM that was misclassified as GAMBLING (Fig. 5f). A counterfactual activation was obtained by transforming the real activation toward WM (Fig. 5g). As in the first example, the counterfactual explanation showed a pattern of brain activation with multiple identifiable peaks (Fig. 5h). In contrast, the difference with the average provided a noisier pattern whose local peaks were difficult to identify (Fig. 5i). These results demonstrated that counterfactual explanation can provide interpretable patterns of brain activations related to individual cases of misclassifications by the DNN classifier.

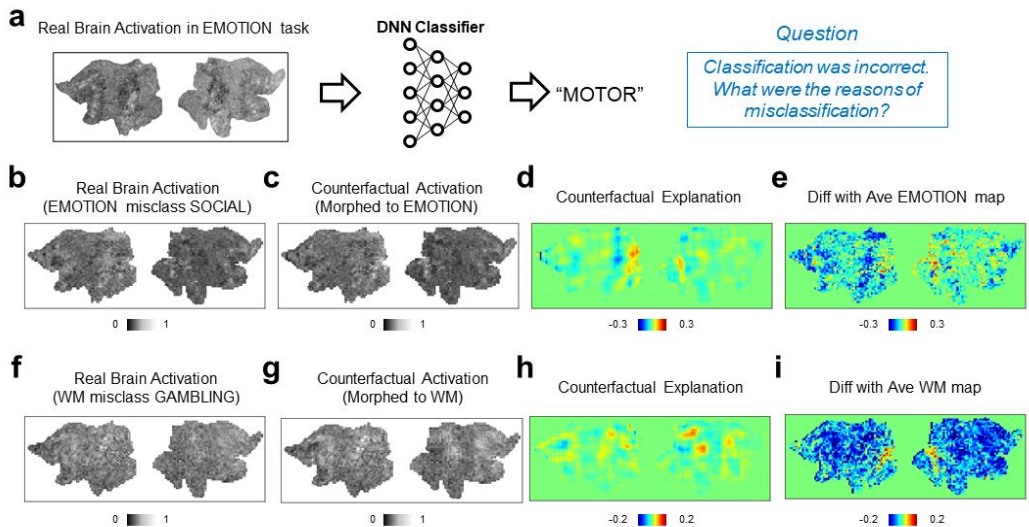

**Figure 5. Counterfactual explanation of incorrect classification. a)** Schematics of the question asked in this analysis. In this example, the DNN classifier incorrectly assigned a "SOCIAL" label to a real brain activation in the EMOTION task. Here, we want to interrogate this incorrect decision. Specifically, we ask a question "why did the classifier (incorrectly) assign EMOTION instead of SOCIAL?". **b-e)** Example of counterfactual explanation. (b) A single brain activation map for EMOTION that was incorrectly classified as SOCIAL by the DNN classifier. (c) A map of counterfactual activation obtained by transforming the map in (a) to SOCIAL. (d) Pixel-by-pixel subtraction of maps in (b) and (a) that serves as counterfactual explanation. This map explains why the map was incorrectly classified as SOCIAL but not EMOTION. (e) Simple difference between the average of real activations in the EMOTION and the single activation map for SOCIAL shown in (a). **f-i)** Another example of counterfactual explanation. Same convention as in (b)-(e). In this case, we asked why brain activations were (incorrectly) classified as GAMBLING but not WM.

Next, we quantitatively assessed the counterfactual explanation of misclassifications. The DNN classifier assigned the correct classes to all the counterfactual activations that are equivalent to additions of the real (misclassified) activations and the counterfactual explanations (40 of 40 misclassified activations in the validation set). To assess the robustness of the results to image perturbation, we conducted the same analysis that we used for the correct classification. In the case of misclassification, the DNN classifier assigned the correct classes in 80% of cases, even when the bottom 80% of the image components modified by CAG were perturbed (Table 3). This result suggests that only a small modification to the misclassified activation was necessary to shift the classifier's decision to the correct class.

As for the control analysis, for each misclassified activation, we calculated the control activation that is the sum of the misclassified activation and the difference of averages of the true class and the incorrectly assigned class. In contrast to counterfactual activations, only two of the control activations were classified as the true classes (Table 3). These results suggest that counterfactual explanation, but not the addition of the difference of average activations, captured the image transformation needed to correct the decisions of the DNN classifier.

| Incorrect Cases | Counterfactual Activations | | | | | | Control |
|---|---|---|---|---|---|---|---|
| | $\alpha = 0$ | $\alpha = 20$ | $\alpha = 40$ | $\alpha = 60$ | $\alpha = 80$ | $\alpha = 100$ | |
| $N_{correct}$ (%) | 40 (100%) | 40 (100%) | 39 (97.5%) | 38 (95.0%) | 24 (60.0%) | 0 (0.0%) | 2 (0.5%) |

**Table 3. Decision of the DNN classifier on counterfactual activations obtained from misclassified brain activations.** Counterfactual activations were obtained from images originally misclassified by the DNN classifier (N = 40). All of the counterfactual activations were correctly classified by the DNN classifier after transformation by CAG (middle column). As for the control analysis, the difference of the average maps for incorrect and correct classes was added to each misclassified image. None of the control images were classified as the targeted class (right column).

*Counterfactual exaggeration revealed subtle image features important for the classifications by DNN*

In addition to counterfactual explanations of correct and incorrect classifications, the deep image generator can perform "counterfactual exaggeration" to enhance and detect subtle image features exploited by DNNs (Singla et al., 2020). In counterfactual exaggeration, an image is iteratively transformed by the generator toward one class. This iterative image transformation enhances subtle image features exploited by DNNs. In a previous work, exaggerated images were used to discover a novel symptom of diabetic macular edema (Narayanaswamy et al., 2020). Inspired by these previous works, we next used CAG in counterfactual exaggeration to detect subtle features of brain activations exploited by the DNN classifier (Fig. 6a). Interestingly, in some cases, iterative application of CAG revealed a texture-like feature in the image (Fig. 6b). Such a texture-like feature was difficult to discern in the original activation (Fig. 6b, left) but became evident as the counterfactual exaggeration was repeatedly applied (Fig. 6b, middle and right). The texture-like feature could be extracted by taking the difference of counterfactual activations with different numbers of iterations (Fig. 6c).

Although the texture-like pattern did not appear in the same way as a real brain activation, it could nevertheless influence the classifier decisions. In fact, it has been suggested that DNNs are biased toward using textures for image classification (Geirhos et al., 2018). To quantitatively examine this point, we added the extracted features to randomly chosen real activations and then examined the resulting activations by the DNN classifier. Figures 6d and 6e show examples of the extracted features and the real activations before and after the addition of the features. Note that differences between the appearance of activations before and after the addition of the features were subtle because the amplitudes of the extracted features were relatively small. Nevertheless, the addition of the extracted features caused the DNN classifier to (mis-)assign the activations the classes to which the exaggerations were targeted (Fig. 6f). Misclassification to the targeted class occurred in $55.0 \pm 26.1\%$ of cases (mean $\pm$ standard deviation; N=12 extracted features; $P < 0.001$, sign rank test; see Methods for details). These results suggest that counterfactual exaggeration assisted by CAG was able to enhance and discover subtle image features that are exploited by the DNN classifier. The texture-like features likely represent image features relevant to adversarial vulnerability of the DNN classifier (Geirhos et al., 2018). Being able to detect and protect against such attacks is critical for future reliable applications of DNN-based brain decoders.

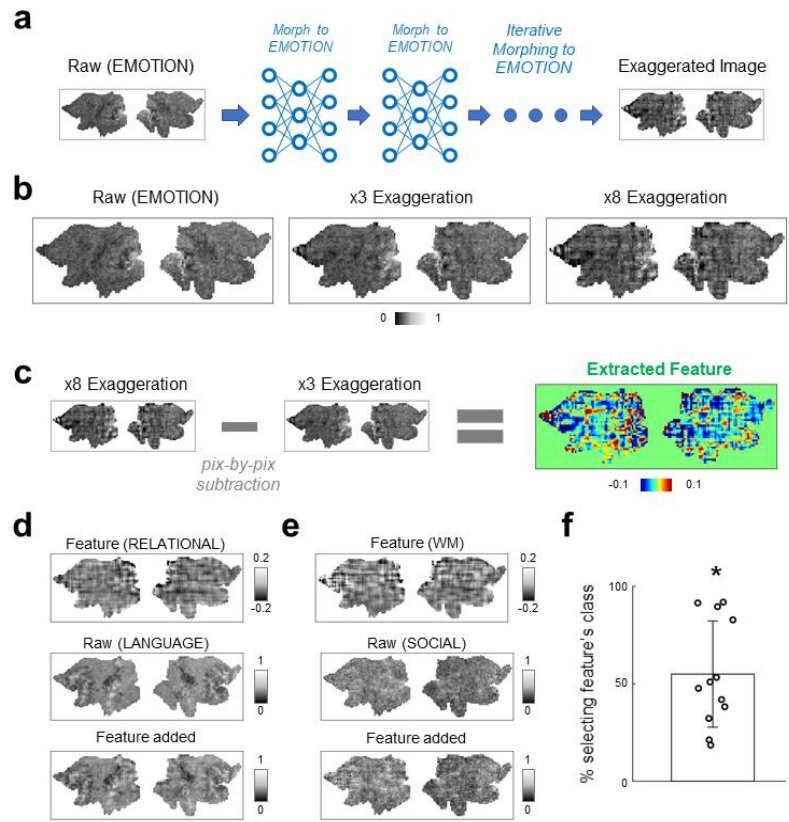

**Figure 6. Counterfactual exaggeration of brain activation. a)** Schematic of counterfactual exaggeration. A brain activation (MOTOR task in this example) was iteratively transformed toward MOTOR by CAG. This iterative transformation accentuates (exaggerates) image features that biases the classifier decision toward MOTOR. **b)** Example of counterfactual exaggeration. A brain activation in the MOTOR task was iteratively transformed toward MOTOR eight times. Images after third (middle) and eighth (right) transformations are shown. **c)** Subtle image feature enhanced by counterfactual exaggeration was isolated by taking the difference of exaggerated images. In this example, differences between exaggerated images in (b) were calculated. The resulting difference image showed a texture-like pattern (right). **d-e)** Examples of texture-like feature extracted by counterfactual exaggeration (top). Bottom panel shows the texture-like patterns added to randomly chosen raw brain activations (middle). **f)** Decisions of the DNN classifier to brain activations with texture-like patterns added. Each dot represents one example texture (N = 12. See Methods for details). Bar graph shows the mean and the standard deviation. The classifier was significantly biased toward the class of texture-like patterns ($P < 0.001$, Wilcoxon's sign rank test). Chance level was one of seven.

There are several limitations in the present study. The training and testing of the DNN classifier and CAG were performed using only the HCP dataset. As more and more neuroimaging datasets become available to the public, researchers are starting to develop DNN classifiers trained on multiple datasets. Though it is beyond the scope of the present study, explaining the DNN classifiers trained on multiple datasets would be an important future research topic. Another limitation is that the present study used spatial down-sampling to enable efficient learning by CAG. This was partly due to limitations in both the computational power and the dataset size. The limitation in the dataset size may be alleviated by using techniques for data augmentation (Shorten and Khoshgoftaar, 2019).

It should also be emphasized that the aim of CAG is not to improve the accuracy of the DNN-classifier but to provide visual explanations for the classifier's decisions. Because CAG can simultaneously take into account information from the entire brain, counterfactual explanation is different from conventional analyses of local activation patterns such as GLM and search light-based multivariate pattern analyses (Chikazoe et al., 2014; Jimura and Poldrack, 2012; Kriegeskorte et al., 2006). This characteristic of CAG is most pronounced in counterfactual exaggerations, where it discovered global texture-like patterns that could effectively bias the classifier's decisions. At present, though, these patterns are unlikely to reflect biologically important activity patterns. Further development of CAG and related techniques would enable the discovery of global activity patterns with biological significance beyond conventional analyses.

*Conclusions*

In this study, we developed CAG, a generative neural network for counterfactual brain activation that can be used to explain individual decision behaviors of DNN-based classifiers. A single CAG could handle multiple classes at the same time and learn mapping between all the pairs of classes. CAG could provide visually intuitive counterfactual explanations for a classifier's correct and incorrect decisions. These counterfactual explanations were quantitatively more effective in explaining the classifier's decision than the controls and were robust against image perturbations. Finally, beyond explaining the decision behaviors, CAG could extract subtle image features in the brain activation that were invisible to the eyes but that were exploited by the DNN classifiers. Together, these results suggest that counterfactual explanation with CAG provides a novel approach to examine and extend current neuroimaging studies using DNNs.

# References

Adadi, A., and Berrada, M. (2018). Peeking Inside the Black-Box: A Survey on Explainable Artificial Intelligence (XAI). Ieee Access *6*, 52138-52160. 10.1109/ACCESS.2018.2870052.

Arjovsky, M., Chintala, S., and Bottou, L. (2017). Wasserstein GAN. arXiv, arxiv:1701.07875.

Barch, D., Burgess, G., Harms, M., Petersen, S., Schlaggar, B., Corbetta, M., Glasser, M., Curtiss, S., Dixit, S., Feldt, C., et al. (2013). Function in the human connectome: Task-fMRI and individual differences in behavior. Neuroimage *80*, 169-189. 10.1016/j.neuroimage.2013.05.033.

Chang, C.-H., Creager, E., Goldenberg, A., and Duvenaud, D. (2019). EXPLAINING IMAGE CLASSIFIERS BY

COUNTERFACTUAL GENERATION. International Conference on Learning Representations (ICLR).

Chattopadhay, A., Sarkar, A., Howlader, P., and Balasubramanian, V. (2018). Grad-CAM plus plus : Generalized Gradient-based Visual Explanations for Deep Convolutional Networks. 2018 Ieee Winter Conference on Applications of Computer Vision (Wacv 2018), 839-847. 10.1109/WACV.2018.00097.

Chikazoe, J., Lee, D.H., Kriegeskorte, N., and Anderson, A.K. (2014). Population coding of affect across stimuli, modalities and individuals. Nat Neurosci *17*, 1114-1122. 10.1038/nn.3749.

Choi, Y., Choi, M., Kim, M., Ha, J.-W., Kim, S., and Choo, J. (2018). StarGAN: Unified Generative Adversarial Networks for Multi-Domain Image-to-Image Translation. IEEE Conference on Computer Vision and Pattern Recognition (CVPR).

Dong, Y., Su, H., Wu, B., Li, Z., Liu, W., Zhang, T., and Zhu, J. (2019). Efficient Decision-based Black-box Adversarial Attacks on Face Recognition. 2019 Ieee/cvf Conference on Computer Vision and Pattern Recognition (Cvpr 2019), 7706-7714. 10.1109/CVPR.2019.00790.

Geirhos, R., Rubisch, P., Michaelis, C., Bethge, M., Wichmann, F.A., and Brendel, W. (2018). ImageNet-trained CNNs are biased towards textures; increasing shape bias increases robustness. International Conference on Learning and Representations (ICLR).

Glasser, M.F., Smith, S.M., Marcus, D.S., Andersson, J.L., Auerbach, E.J., Behrens, T.E., Coalson, T.S., Harms, M.P., Jenkinson, M., Moeller, S., et al. (2016). The Human Connectome Project's neuroimaging approach. Nat Neurosci *19*, 1175-1187. 10.1038/nn.4361.

Goodkind, M.S., Sollberger, M., Gyurak, A., Rosen, H.J., Rankin, K.P., Miller, B., and Levenson, R. (2012). Tracking emotional valence: the role of the orbitofrontal cortex. Hum Brain Mapp *33*, 753-762. 10.1002/hbm.21251.

Goyal, Y., Wu, Z., Ernst, J., Batra, D., Parikh, D., and Lee, S. (2019). Counterfactual Visual Explanations. arXiv. arXiv:1904.07451v2.

Jimura, K., and Poldrack, R.A. (2012). Analyses of regional-average activation and multivoxel pattern information tell complementary stories. Neuropsychologia *50*, 544-552. 10.1016/j.neuropsychologia.2011.11.007.

Kim, T., Cha, M., Kim, H., Lee, J., Kim, J., Precup, D., and Teh, Y. (2017). Learning to Discover Cross-Domain Relations with Generative Adversarial Networks. International Conference on Machine Learning, Vol 70 *70*.

Kingma, D., and Ba, J. (2014). Adam: a method for stochastic optimization. arXiv, arxiv:1412.6980.

Kriegeskorte, N., Goebel, R., and Bandettini, P. (2006). Information-based functional brain mapping. Proc Natl Acad Sci U S A *103*, 3863-3868. 10.1073/pnas.0600244103.

LeCun, Y., Bengio, Y., and Hinton, G. (2015). Deep learning. Nature *521*, 436-444. 10.1038/nature14539.

Liu, S., Kailkhura, B., Loveland, D., and Han, Y. (2019). Generative Counterfactual Introspection for Explainable Deep Learning. 2019 7th Ieee Global Conference on Signal and Information Processing (Ieee Globalsip).

Mertes, S., Huber, T., Weitz, K., Heimerl, A., and Andre, E. (2020). GANterfactual - Counterfactual Explanation for Medical Non-Experts using Generative Adversarial Learning. arXiv, arrive:2012.11905v11903.

Narayanaswamy, A., Venugopalan, S., Webster, D.R., Peng, L., Corrado, G.S., Ruamviboonsuk, P., Bavishi, P., Brenner, M., Nelson, P.C., and Varadarajan, A.V. (2020). Scientific Discovery by Generating Counterfactuals using Image Translation. International Conference on Medical Image Computing and Computer-Assisted Intervention (MICCAI).

Pan, S., and Yang, Q. (2010). A Survey on Transfer Learning. Ieee Transactions on Knowledge and Data Engineering *22*, 1345-1359. 10.1109/TKDE.2009.191.

Pawlowski, N., Castro, D.C., and Glocker, B. (2020). Deep Structural Causal Models

for Tractable Counterfactual Inference. Conference on Neural Information Processing Systems (NeurIPS).

Poldrack, R.A. (2006). Can cognitive processes be inferred from neuroimaging data? Trends Cogn Sci *10*, 59-63. 10.1016/j.tics.2005.12.004.

Rolls, E.T., Cheng, W., and Feng, J. (2020). The orbitofrontal cortex: reward, emotion and depression. Brain Commun *2*, fcaa196. 10.1093/braincomms/fcaa196.

Selvaraju, R., Cogswell, M., Das, A., Vedantam, R., Parikh, D., and Batra, D. (2020). Grad-CAM: Visual Explanations from Deep Networks via Gradient-Based Localization. International Journal of Computer Vision *128*, 336-359. 10.1007/s11263-019-01228-7.

Shorten, C., and Khoshgoftaar, T. (2019). A survey on Image Data Augmentation for Deep Learning. Journal of Big Data *6*, ARTN 60. 10.1186/s40537-019-0197-0.

Simonyan, K., and Zisserman, A. (2015). Very Deep Convolutional Networks for Large-Scale Image Recognition. International Conference on Learning Representations (ICLR).

Singla, S., Pollack, B., Chen, J., and Batmanghelich, K. (2020). Explanation by Progressive Exaggeration. International Conference on Learning Representations (ICLR).

Tsumura, K., Kosugi, K., Hattori, Y., Aoki, R., Takeda, M., Chikazoe, J., Nakahara, K., and Jimura, K. (2021). Reversible Fronto-occipitotemporal Signaling Complements Task Encoding and Switching under

Ambiguous Cues. Cereb Cortex. 10.1093/cercor/bhab324.

Van Essen, D.C., Smith, S.M., Barch, D.M., Behrens, T.E., Yacoub, E., Ugurbil, K., and Consortium, W.-M.H. (2013). The WU-Minn Human Connectome Project: an overview. Neuroimage *80*, 62-79. 10.1016/j.neuroimage.2013.05.041.

Wang, P., and Vasconcelos, N. (2020). SCOUT: Self-Aware Discriminant Counterfactual Explanations. Proceedings of the IEEE/CVF Conference on Computer Vision and Pattern Recognition (CVPR).

Wang, X., Liang, X., Jiang, Z., Nguchu, B.A., Zhou, Y., Wang, Y., Wang, H., Li, Y., Zhu, Y., Wu, F., et al. (2020). Decoding and mapping task states of the human brain via deep learning. Hum Brain Mapp *41*, 1505-1519. 10.1002/hbm.24891.

White, S.F., Adalio, C., Nolan, Z.T., Yang, J., Martin, A., and Blair, J.R. (2014). The amygdala's response to face and emotional information and potential category-specific modulation of temporal cortex as a function of emotion. Front Hum Neurosci *8*, 714. 10.3389/fnhum.2014.00714.

Yamins, D.L., Hong, H., Cadieu, C.F., Solomon, E.A., Seibert, D., and DiCarlo, J.J. (2014). Performance-optimized hierarchical models predict neural responses in higher visual cortex. Proc Natl Acad Sci U S A *111*, 8619-8624. 10.1073/pnas.1403112111.

Yarkoni, T., Poldrack, R.A., Nichols, T.E., Van Essen, D.C., and Wager, T.D. (2011). Large-scale automated synthesis of human functional neuroimaging data. Nat Methods *8*, 665-670. 10.1038/nmeth.1635.

Zhao, Y. (2020). Fast Real-time Counterfactual Explanations. arXiv. arXiv:2007.05684v2.

Zhu, J., Park, T., Isola, P., and Efros, A. (2017). Unpaired Image-to-Image Translation using Cycle-Consistent Adversarial Networks. 2017 Ieee International Conference on Computer Vision (Iccv), 2242-2251. 10.1109/ICCV.2017.244.

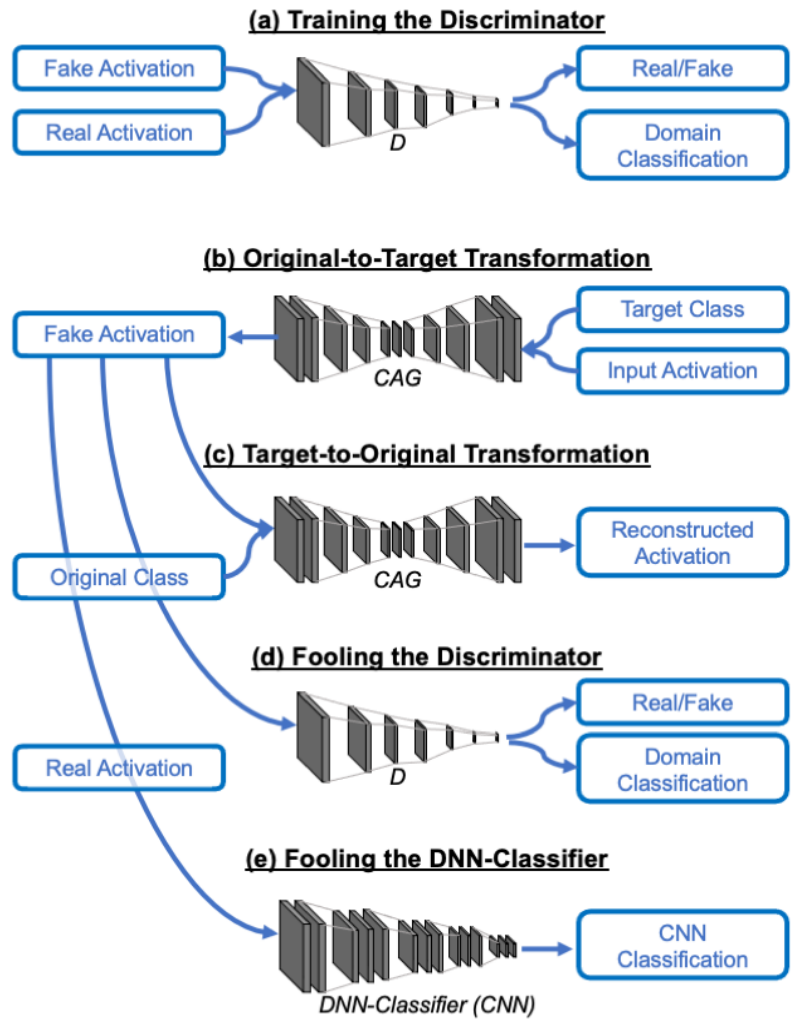

## Supplementary Information

**Figure S1. Overall architecture of Discriminator/CAG training.** We adapted the overall design of the StarGAN(Choi *et al.*, 2018). The architecture includes two trainable modules, Discriminator (D) and CAG. Note that DNN-classifier is not trainable. (a) D learns to discriminate between real and fake activations and to classify the real activation to its corresponding domain. (b) CAG takes a real activation and a target domain label (*i.e.* target class) to generate a fake activation. (c) CAG tries to reconstruct the original activation from the fake activation and the domain label of the original activation. (d) CAG tries to generate a fake activation that is indistinguishable from real activations and classifiable as target domain by D. We also added a process

"Fooling the DNN-Classifier" (e) to force CAG to generate a fake activation that is classifiable as target domain by the DNN-classifier.

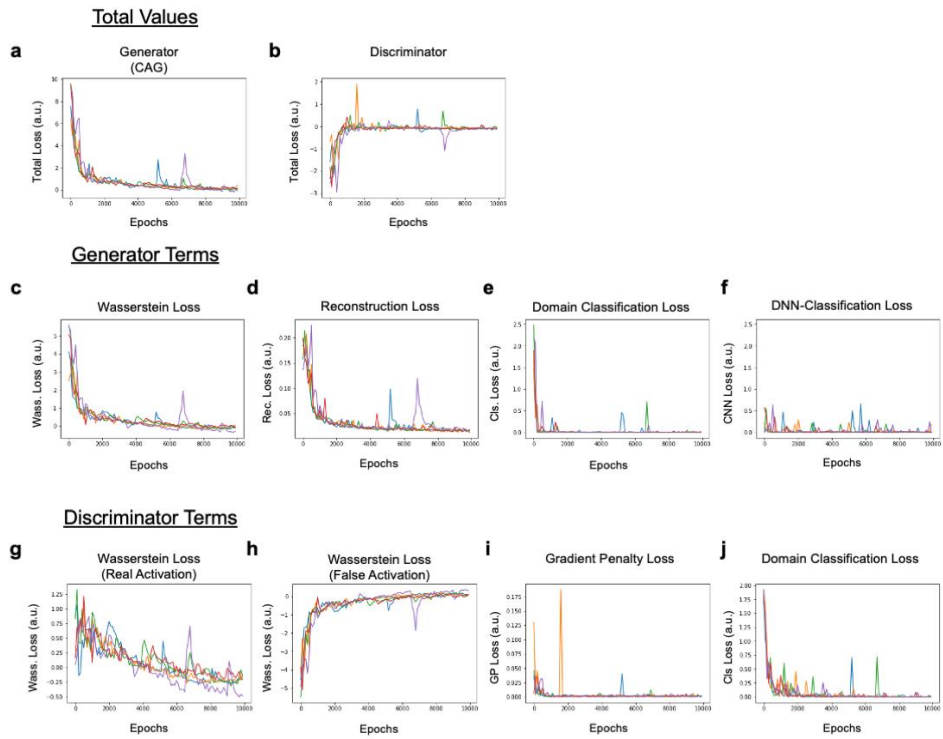

**Figure S2 Time courses of key terms related to the Discriminator/CAG training. a)** Total Generator Loss for CAG (same as Fig.3b). **b)** Total Discriminator Loss for Discriminator. **c-f)** Decomposed Generator Loss. **g-j)** Decomposed Discriminator Loss. Five independent replicates are shown in different colors.

**Table S1. Profile of the DNN classifier (Confusion Matrix)** Same as Fig.2d but exact numbers for proportion correct are shown)

|  |  | Output Class | | | | | | |
| --- | --- | --- | --- | --- | --- | --- | --- | --- |
|  |  | EMOTION | GAMBLING | LANGUAGE | MOTOR | RELATIONAL | SOCIAL | WM |
| Input Class | EMOTION | 0.986 | 0 | 0 | 0 | 0 | 0.014 | 0 |
|  | GAMBLING | 0 | 0.877 | 0 | 0 | 0.068 | 0.055 | 0 |
|  | LANGUAGE | 0 | 0 | 0.986 | 0 | 0 | 0.014 | 0 |
|  | MOTOR | 0 | 0 | 0 | 1 | 0 | 0 | 0 |
|  | RELATIONAL | 0 | 0.095 | 0 | 0.027 | 0.851 | 0.014 | 0.014 |
|  | SOCIAL | 0.014 | 0.027 | 0 | 0 | 0.027 | 0.932 | 0 |
|  | WM | 0 | 0.107 | 0 | 0.027 | 0.04 | 0 | 0.827 |